\documentclass[aps,prl,twocolumn,superscriptaddress]{revtex4-1}
\usepackage{graphicx}

\begin{document}

\title{Anisotropic Intermittency of Magnetohydrodynamic Turbulence}

\author{K.T. Osman}
\email{k.t.osman@warwick.ac.uk}
\affiliation{Centre for Fusion, Space and Astrophysics; University of Warwick, Coventry, CV4 7AL, UK}

\author{K.H. Kiyani}
\affiliation{Laboratoire de Physique des Plasmas, \'Ecole Polytechnique, Route de Saclay, 91128 Palaiseau, France}
\affiliation{Centre for Fusion, Space and Astrophysics; University of Warwick, Coventry, CV4 7AL, UK}

\author{S.C. Chapman}
\affiliation{Centre for Fusion, Space and Astrophysics; University of Warwick, Coventry, CV4 7AL, UK}
\affiliation{Department of Mathematics and Statistics, University of Troms\o, N-9037 Troms\o, Norway}
\affiliation{Max Planck Institute for the Physics of Complex Systems, 01187 Dresden, Germany}

\author{B. Hnat}
\affiliation{Centre for Fusion, Space and Astrophysics; University of Warwick, Coventry, CV4 7AL, UK}

\date{\today}

\begin{abstract}

A higher-order multiscale analysis of spatial anisotropy in inertial range magnetohydrodynamic turbulence is presented using measurements from the STEREO spacecraft in fast ambient solar wind. We show for the first time that, when measuring parallel to the local magnetic field direction, the full statistical signature of the magnetic and Els\"asser field fluctuations is that of a non-Gaussian globally scale-invariant process. This is distinct from the classic multi-exponent statistics observed when the local magnetic field is perpendicular to the flow direction. These observations are interpreted as evidence for the weakness, or absence, of a parallel magnetofluid turbulence energy cascade. As such, these results present strong observational constraints on the statistical nature of intermittency in turbulent plasmas.

\end{abstract}

\pacs{}

\maketitle

\textit{Introduction}.---Turbulence is a universal fluid phenomenon that generates intermittent fluctuations \citep{BrunoCarbone13}. The solar wind provides an ideal laboratory for the \textit{in-situ} study of plasma turbulence, wherein intermittent fluctuations have been analyzed in considerable detail. These have been linked to non-uniform plasma heating \citep{OsmanEA11a,OsmanEA12a,WuEA13} and enhanced turbulent dissipation \citep{WanEA12,KarimabadiEA13,TenBargeHowes13}. In addition, there is evidence to suggest increased alpha particle \citep{PerroneEA13}, proton \citep{ServidioEA12} and electron \citep{HaynesEA13} temperature anisotropies are associated with intermittent structures. The same structures can cause particle velocity distribution functions to deviate from local thermal equilibrium \citep{GrecoEA12} and have been preferentially found in plasma unstable to microinstabilities \citep{OsmanEA12b}. A subset of non-Gaussian intermittent structures correspond to active magnetic reconnection sites \citep{ServidioEA11} which can in turn generate fluctuations that exhibit the hallmarks of intermittency \citep{LeonardisEA13}. These spatial structures may also be related to trapping boundaries that delineate dropouts of energetic particle flux as seen in solar energetic particle data \citep{RuffoloEA03}. Indeed, recent work suggests that these structures contribute to the acceleration and transport of interplanetary suprathermal particles \citep{TesseinEA13}. Intermittent fluctuations are the reason why turbulence can enhance the transport of particles, heat, momentum and current in laboratory plasmas. The intermittent structures in plasmas share striking similarities with fluctuations found in turbulent neutral fluids \citep{Frisch95, SreenivasanAntonia97}. Therefore, quantifying intermittency is central to understanding and interpreting a large body of observations in turbulent systems.

Intermittency lies at the heart of turbulence theory. The classical signatures of intermittency in both neutral fluid and MHD turbulence are a non-Gaussian probability distribution function (PDF) of fluctuations and multifractal scaling in the higher order statistics \citep{BrunoCarbone13}. However, non-fluid phenomenology such as the kinetic range of plasma turbulence can have a monofractal scaling \citep{KiyaniEA09}. These different scaling types imply different underlying physics. We will quantify both the non-Gaussian behavior of fluctuations and their statistical scaling for the fast quiet turbulent solar wind. Intermittency is related to the emergence of small-scale coherent structures that are responsible for enhanced dissipation. Hence, the most fundamental approach to the study of intermittency is to examine the dissipation rate PDF. However, the Kolmogorov refined similarity hypothesis \citep[hereafter KRSH;][]{Kolmogorov62,Oboukhov62} allows local averages of the dissipation rate to be related to increments of the velocity field calculated on different spatial scales, $r$. The PDF of velocity increments is then linked to intermittency \citep{AnselmetEA84}, where departures from a normal distribution occur on small spatial scales while large scale features are uncorrelated and converge towards a Gaussian distribution. This non-Gaussian behavior is also observed in the turbulent solar wind magnetic field \citep{Sorriso-ValvoEA99}. A method to quantify intermittency is based on computing a sequence of $m$th order moments of the magnetic or velocity field increments. For an increment scale $r$, the moments have power-law scalings, $\propto r^{\zeta}$, where the exponents $\zeta$ depend on the moment order $m$. Here the physical meaning lies in the sensitivity of higher order moments to concentrations of dissipation and, from KRSH, to large increments. The behavior of these exponents is also connected to known fractal and multifractal models \citep{FrischEA78,SheLeveque94,PolitanoPouquet95}. This Letter presents novel observational results from a higher-order analysis that examines the statistical properties of MHD turbulence in the spatially anisotropic solar wind. We find for the first time that these statistical properties depend on the angle of the local magnetic field direction to the (radial) solar wind flow. This provides strong constraints on the physics and phenomenology of inertial range turbulence in collisionless plasmas.

The presence of a magnetic field in plasma turbulence breaks the isotropy found in hydrodynamics and orders the fluctuations \citep{HorburyEA05}. In the solar wind, fluctuation components parallel and transverse to the background magnetic field display differences in dynamics and statistics \citep{ChapmanHnat07}. However, this `variance' anisotropy is not reflected in the higher-order moments and both components display a multifractal intermittent scaling \citep{KiyaniEA13}. The distribution of energy over the full three-dimensional space of wavevectors is also anisotropic \citep{OughtonEA94}. This spatial anisotropy has been observed in second order statistics such as the power spectral density \citep{OsmanHorbury09b} and correlation function \citep{MatthaeusEA90,OsmanHorbury07,OsmanHorbury09a}. It has also been found in third order statistics \citep{OsmanEA11b}. A higher-order analysis of wavevector anisotropy would provide a direct test of theoretical predictions regarding the statistical nature of intermittency and, more broadly, the phenomenology of the turbulent cascade. However, an investigation into the wavevector anisotropy of intermittent fluctuations has not, to the best of our knowledge, been conducted.

\textit{Analysis}.---We use 8 Hz magnetic field measurements from the IMPACT instrument \citep{AcunaEA08,LuhmannEA08} and 1 min resolution proton plasma data from the PLASTIC instrument \citep{GalvinEA08} onboard the two STEREO spacecraft in the ecliptic. The solar wind intervals used here are all in high-speed streams and contain no sector crossings. These are listed in Table 1 and are identical to those used by \citep{Podesta09}. It has been suggested \citep[e.g.][]{ChapmanHnat07,HorburyEA08} that a local scale-dependent mean magnetic field and associated scale-dependent fluctuations, rather than a large-scale global field \citep{MatthaeusEA12}, should be used in anisotropy studies of plasma turbulence. Hence, we use the undecimated discrete wavelet transform (UDWT) method described in \citep{KiyaniEA13} to decompose the magnetic field into a local scale-dependent background and fluctuations, $\bar{\mathbf{B}}(t,f)$ and $\delta\mathbf{B}(t,f)$, where $f$ explicitly shows the frequency or scale dependence. These fluctuations are binned according to the angle of the local magnetic field direction to the (radial) flow, $\theta_{VB}$. Here we focus on fluctuations in the $\theta_{VB} = 0^{\circ}$--$10^{\circ}$ and $80^{\circ}$--$90^{\circ}$ bins, which correspond to wavenumbers using Taylor's hypothesis \citep{Taylor38} that are respectively near field-parallel $\delta\mathbf{B}(k_{\parallel})$ and near field-perpendicular $\delta\mathbf{B}(k_{\perp})$.

\begin{table}[h]
\caption{List of all the high-speed streams in the ecliptic plane that are analyzed. Here S/C represents spacecraft, where STA is STEREO A and STB is STEREO B.}
\begin{tabular}{l  c  c  c  c  c}
\hline \hline
No. & Year & Start & End & Days & S/C \\
\hline
1 & 2007 & 28 Apr 00:00 & 01 May 00:00 & 3 & STB \\
2 & 2007 & 25 May 00:00 & 28 May 02:39 & 3.11 & STB \\
3 & 2007 & 27 Aug 12:00 & 30 Aug 12:00 & 3 & STB \\
4 & 2007 & 15 Nov 00:00 & 18 Nov 00:00 & 3 & STA \\
5 & 2008 & 08 Jan 00:00 & 11 Jan 00:00 & 3 & STA \\
6 & 2008 & 13 Feb 00:00 & 18 Feb 00:00 & 5 & STA \\
7 & 2008 & 08 Mar 00:00 & 11 Mar 00:00 & 3 & STB \\ 
8 & 2008 & 04 Apr 00:00 & 08 Apr 00:00 & 4 & STB \\
9 & 2008 & 02 May 12:00 & 06 May 00:00 & 3.5 & STB \\
\hline
\end{tabular}
\end{table}

\textit{Results}.---We present a detailed analysis of interval 8 listed in Table 1, which is typical of the stationary fast solar wind intervals used in this study. The power spectral density (PSD) is independent of the azimuthal angle about the local magnetic field for all relevant spacecraft frame frequencies \citep{Podesta09}. Hence, the spacecraft frame wavelet PSD depends only on the angle of the magnetic field to the flow direction:
\begin{equation}
PSD(f,\theta_{VB}) = \frac{2\Delta}{N}\sum\limits_{j=1}^{N} \delta B^{2}(t_{j},f,\theta_{VB})
\end{equation} 
where $\delta B(t_{j},f,\theta_{VB})$ is the magnitude of the trace fluctuations at time $t_{j}$ and frequency $f$, $\Delta$ is the sampling time between consecutive measurements, and $N$ is the sample size at each frequency. Figure 1 shows the PSD for two angular bins, $\theta_{VB} = 0^{\circ}$--$10^{\circ}$ and $80^{\circ}$--$90^{\circ}$, which correspond to wavevectors roughly parallel and perpendicular to the local field respectively. These are both well described by power laws. The power levels are lower and the spectral slope is steeper for $\delta\mathbf{B}(k_{\parallel})$ compared to $\delta\mathbf{B}(k_{\perp})$, which is consistent with previous studies \citep[e.g.][]{HorburyEA08,Podesta09}. Hence, the statistical behavior of fluctuations in wavevectors at large angles to the magnetic field would have dominated all previous estimates of the inertial range intermittency, since these contain the most power. The dashed vertical lines define the range of timescales used in the higher order analysis. 

\begin{figure}[h]
\includegraphics[width=8.5cm]{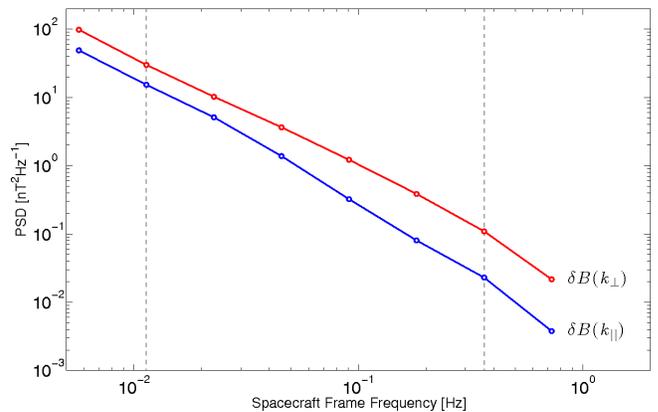}
\caption{PSD of the trace magnetic field fluctuations  for the angular bins $\theta_{VB} = 0^{\circ}$--$10^{\circ}$ (squares) and $80^{\circ}$--$90^{\circ}$ (circles).}
\label{Fig:PSD}
\end{figure}

In order to determine the higher order scaling of fluctuations for different $\theta_{VB}$, we compute the absolute moments of the magnetic field increments, $\delta B(t,\tau) = B(t+\tau) - B(t)$. The $m$th order wavelet structure function is given by:
\begin{equation}
S^{m}(\tau,\theta_{VB}) = \frac{1}{N}\sum\limits_{j=1}^{N} \left| \frac{\delta B(t_{j},\tau,\theta_{VB})}{\sqrt{\tau}} \right|^{m}
\end{equation}
where $\tau = 2^{i}\Delta: i = \{0, 1, 2, 3, \dots\}$ is the dyadic timescale parameter related to the central frequency $f$. Note that wavelets change the regular expressions for structure functions \citep{KiyaniEA13}. The higher-order structure functions increasingly capture the more intermittent fluctuations. In hydrodynamics, these large fluctuations represent the spatial gradients responsible for dissipating the turbulent cascade energy. However, there is growing evidence to suggest these intermittent structures are also associated with nonuniform heating \citep{OsmanEA11a,OsmanEA12a,WuEA13} and increased temperature anisotropy \citep{ServidioEA12,OsmanEA12b,KarimabadiEA13} in plasma turbulence. 

Here the focus will be on the scaling behavior of structure functions, where scale invariance is indicated by:
\begin{equation}
S^{m}(\tau) \propto \tau^{\zeta(m)}
\end{equation}
and $\zeta(m)$ are the scaling exponents. Figure 2 shows the scaling exponents for both the $\theta_{VB} = 0^{\circ}$--$10^{\circ}$ and $80^{\circ}$--$90^{\circ}$ angular bins. The higher order scaling of the magnetic field-parallel and perpendicular fluctuations are distinct; this is a novel result. The $\theta_{VB} = 80^{\circ}$--$90^{\circ}$ fluctuations have a nonlinear $\zeta(m)$ that indicates a multiexponent scaling, which is characteristic of hydrodynamic turbulence \citep{Frisch95} and solar wind turbulence at MHD scales \citep{BrunoCarbone13}. In contrast, the $\theta_{VB} = 0^{\circ}$--$10^{\circ}$ fluctuations are characterized by a linear $\zeta(m) = Hm$ with a single exponent $H$, which indicates monoscaling. The errors on $\zeta(m)$ shown in Fig. 2 were obtained from the sum of the regression error when using Eq. 3, and from variations in $\zeta(m)$ that resulted from repeating the same regression over a subinterval of the original scaling range \citep{KiyaniEA06}. In order to confirm the robustness of this result, the analysis was repeated for all nine intervals listed in Table 1 and the same $\theta_{VB}$ dependent intermittency was obtained.

\begin{figure}[h]
\includegraphics[width=8.5cm]{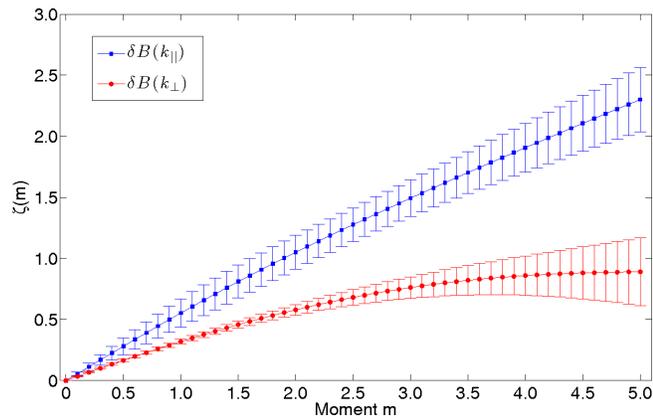}
\caption{The scaling exponents $\zeta(m)$ for the trace magnetic field fluctuations. For fluctuations with $\theta_{VB} = 0^{\circ}$--$10^{\circ}$, there is a linear relationship on this plot that indicates fractal scaling. There is a distinct nonlinear (concave) behavior for fluctuations with $\theta_{VB} = 80^{\circ}$--$90^{\circ}$, which indicates a multifractal.}
\label{Fig:Bscale}
\end{figure}

The statistical analysis is completed by examining scale-by-scale the PDF for the $\theta_{VB} = 0^{\circ}$--$10^{\circ}$ fluctuations. A component of the trace magnetic field fluctuations is selected in order to show any symmetric or asymmetric behavior in the fluctuations. Here we use one of the components transverse to the local field, although the behavior is identical for all three vector components of the fluctuations. Since global scale invariance of the structure functions implies that the PDF of the increments at a scale $\tau$ should collapse onto a unique scaling function $P_{s}$, we use the self-affine scaling operation $P_{s}(\delta B\sigma^{-1}) = \sigma P(\delta B,\tau)$ to rescale the fluctuations by their standard deviation. Figure 3 shows PDFs corresponding to $\tau = \{2, 4, 8, 16, 32\}$ s that are rescaled and overlaid, where the central $\tau$ is plotted in red and shows the associated errors on the PDFs. There is an excellent collapse onto a single curve, although the largest events in the tails of the distribution are not statistically well-sampled as indicated by the large errors, which is an unavoidable consequence of heavy-tailed distributions. In addition, a fitted Gaussian distribution illustrates the highly non-Gaussian nature of the PDF tails and reflects the presence of rare large amplitude fluctuations.

\begin{figure}[h]
\includegraphics[width=8.5cm]{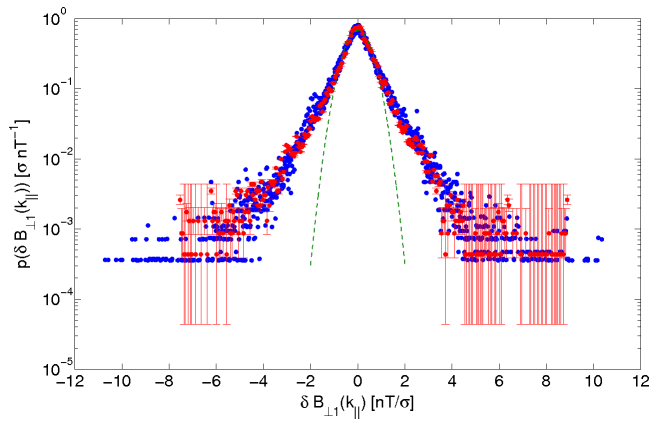}
\caption{PDFs of the $\theta_{VB} = 0^{\circ}$--$10^{\circ}$ trace magnetic field fluctuations rescaled using $P_{s}(\delta B\sigma^{-1}) = \sigma P(\delta B,\tau)$. A Gaussian fit applied to the data (dashed curve) illustrates the heavy-tailed non-Gaussian nature of the rescaled PDF.} 
\label{Fig:PDF}
\end{figure}

While the higher-order analysis has focused exclusively on magnetic field fluctuations, it is the total energy (magnetic and velocity) that is cascaded from large to small scales by plasma turbulence. Therefore, it is instructive to examine Els\"asser fluctuations, $\delta\mathbf{z}^{\pm} = \delta\mathbf{V} \pm \delta\mathbf{B}$, since dynamic couplings produce structure in both magnetic and velocity fields. Here the magnetic field has been normalized to Alfv\'en velocity units, $\delta\mathbf{B}/\sqrt{\mu_{0}m_{p}n_{p}}$, and the fluctuations have been sector rectified such that $\delta\mathbf{z}^{-}$ is sunward and $\delta\mathbf{z}^{+}$ is antisunward. Figure 4 shows the scaling exponents for the antisunward Els\"asser variable in both the $\theta_{VB} = 0^{\circ}$--$10^{\circ}$ and $80^{\circ}$--$90^{\circ}$ angular bins. The higher-order scaling of the fluctuations in both these bins is similar to those in Fig. 2 for the magnetic field fluctuations. The $\theta_{VB} = 80^{\circ}$--$90^{\circ}$ Els\"asser fluctuations have a nonlinear $\zeta(m)$, which is typical of MHD scale solar wind turbulence. This behavior is also associated with the energy dissipation intensity being distributed on a spatial multifractal \citep{BrunoCarbone13}. The $\theta_{VB} = 0^{\circ}$--$10^{\circ}$ fluctuations have a linear $\zeta(m)$, which is characteristic of global scale invariance. In theories of turbulence, this scaling behavior is associated with the energy dissipation intensity being distributed on a fractal. This analysis was repeated for the sunward and antisunward Els\"asser fluctuations in all nine intervals listed in Table 1 and the same $\theta_{VB}$ dependent intermittent scaling was obtained. However, the sunward fluctuations have greater associated errors since these are a minority and contain the least power \citep{GogoEA12}.

\begin{figure}[h]
\includegraphics[width=8.5cm]{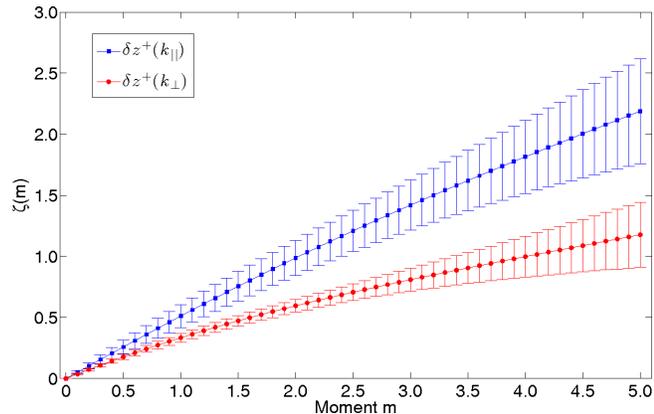}
\caption{The scaling exponents $\zeta(m)$ for the trace antisunward Els\"asser field fluctuations. For fluctuations with $\theta_{VB} = 0^{\circ}$--$10^{\circ}$, there is a linear relationship on this plot that indicates fractal scaling. There is a distinct nonlinear (concave) behavior for fluctuations with $\theta_{VB} = 80^{\circ}$--$90^{\circ}$, which indicates a multifractal scaling. This anisotropic scaling is identical to that observed with the magnetic field fluctuations in Fig. 2.} 
\label{Fig:Zscale}
\end{figure}

\textit{Disscussion}.---We have presented the first direct observation that higher-order scaling of the magnetic and Els\"asser field fluctuations depends on the angle of the local magnetic field direction to the (radial) flow. In fluctuations with wavevectors parallel to the local magnetic field direction, global scale invariance is a robust feature of inertial range collisionless plasma turbulence in the fast ambient solar wind. This is distinct from the multifractal scaling that is characteristic of neutral fluid turbulence and MHD fluctuations with wavevectors perpendicular to the local field. These properties must be included in any successful theory that attempts to explain inertial range intermittency.

A process that has multifractal properties generates fluctuations through a multiplicative sequence such as an energy cascade of eddies in turbulent flows, while monofractal processes generate fluctuations through additive sequences. Therefore, the solar wind MHD turbulence cascade proceeds from smaller to larger wavenumbers that are mainly perpendicular to the local magnetic field direction. The presence of monoscaling parallel to the local field indicates that the cascade in this direction does not proceed in the well understood classic fluid turbulence manner. It could be that kinetic physics is important in the parallel cascade even on what is typically considered MHD scales, and thus behavior associated with kinetic range turbulence such as monoscaling \citep{KiyaniEA09} is observed. Alternatively, the monofractal scaling may be evidence for a weak turbulent cascade. This is consistent with incompressible 3D MHD simulations that found weak (rapid) spectral transfer into wavevectors parallel (perpendicular) to the mean magnetic field \citep{OughtonEA94}. This can be understood in terms of resonant three-wave interactions \citep{ShebalinEA83}. A weak parallel cascade interpretation would also be consistent with several theories and models of collisionless plasma turbulence \citep{OughtonMatthaeus05} such as reduced MHD \citep[e.g.][]{Montgomery82}, `critical balance' \citep{GoldreichSridhar95} and gyrokinetics \citep{SchekochihinEA09}. However, while our results imply that the parallel and perpendicular wavevector cascades proceed with different physics, further work is required to determine the exact nature of these differences.

The present analysis applies to intermittent turbulence exclusively in fast ambient solar wind, and further investigation is required to determine whether the phenomenology of spatially anisotropic intermittency is universal. Hence, similar studies will be conducted in different solar wind streams and plasma environments, such as planetary shocks and magnetospheres, with the aim of reproducing the current results. In addition, work has already begun on investigating the presence of similar spatial anisotropy in the higher-order multiscale analysis of dissipation range turbulence, where a dominant monoscaling has already been observed \citep{KiyaniEA09}.

This research is supported by UK STFC, EPSRC and EU Turboplasmas project (Marie Curie FP7 PIRSES-2010-269297). The authors acknowledge useful conversations with S. Oughton, W. H. Matthaeus and M. Wan.

\end{document}